\documentclass{PoS}

\PoS{PoS(LAT2005)278}

\newcommand{\be}{\begin{equation}}
\newcommand{\ee}{\end{equation}}
\newcommand{\ba}{\begin{eqnarray}}
\newcommand{\ea}{\end{eqnarray}}
\newcommand{\bi}{\begin{itemize}}
\newcommand{\ei}{\end{itemize}}

\newcommand{\Lp}{L_\perp}

\newcommand{\nn}{\nonumber \\}
\newcommand{\<}{\langle} 
\renewcommand{\>}{\rangle} 
\newcommand{\eq}{Eq.~}

\newcommand{\la}{\label}

\title{Kosterlitz-Thouless transition on the worldsheet of the QCD string}

\ShortTitle{Kosterlitz-Thouless transition on the worldsheet of the QCD string}

\author{\speaker{Harvey B. Meyer} \\ DESY \\ Platanenallee 6\\ D-15738 Zeuthen \\
        E-mail: \email{harvey.meyer@desy.de}}

\abstract{We investigate the properties of the QCD string in
the Euclidean SU($N$) pure gauge theory when the
space-time dimensions transverse to it are periodic.
We propose a generalisation of the L\"uscher-Weisz
effective string action for the flux-tube energy levels at finite ${\Lp}$.
As the size of one transverse dimension is varied,
we predict a Kosterlitz-Thouless transition
of the worldsheet field theory at $\sigma(\Lp)\Lp^2\simeq1/8\pi$
driven by vortices, after which the periodic component of the worldsheet
displacement vector develops a mass gap and the
effective central charge drops by one unit.
The universal  properties of the transition are emphasized.}

\FullConference{XXIIIrd International Symposium on Lattice Field Theory\\

                 25-30 July 2005\\

                 Trinity College, Dublin, Ireland}

\begin{document}

\section{Introduction}
Our aim is to investigate the effect on the spectrum of the QCD string
(for a review, see~\cite{kuti_talk}) of periodic transverse dimensions
(\cite{Meyer:2005px}; see~\cite{Caselle:1993cb} for earlier work on the subject).
On the one hand it is important to know the magnitude of the finite-size effects 
on the string spectrum in lattice simulations. By the same token it is hoped that 
one might exploit these effects to learn more about the nature of the 
`flux-tube' degrees of freedom.

Our starting point will be the effective theory 
proposed by L\"uscher and Weisz~\cite{lw04}
for the Polyakov loop correlator in $D=3$ and 4 SU($N$) 
pure gauge theories in infinite transverse dimensions.
%%%%%%%%%%%%%%%%%%%%%%%%%%%%%%%%%%%%%%%%%%%%%%
\section{The L\"uscher-Weisz effective theory for the QCD string\la{sec:eff_th}}
%%%%%%%%%%%%%%%%%%%%%%%%%%%%%%%%%%%%%%%%%%%%%%
This  effective theory~\cite{lw04}  is in principle capable of
predicting the splittings (and the degeneracies)
between low-lying energy levels of the flux-tube,
once a few low-energy constants have been determined.
So it is a low-energy effective theory with a finite
UV-cutoff set by the string tension.
Indeed, at energies of that order, once expects the internal
degrees of freedom of the string to become excited.

The partition function is defined by 
an action living on the worldsheet $R\times L$ of
the string. The connection with the gauge theory
observables is made by equating the Polyakov loop correlation function
$\< P_t(0)^*P_t(x)\>$  with this partition function, 
where $|\vec x|=R$ and the length of the loops is $L$.
For $L\gg R$ the correlator is given
by $e^{-V(R)L}$, where $V(R)$ is interpreted as the  
potential between two static quarks.
The partition function corresponds to
a non-renormalisable Euclidean quantum field
theory in (1+1) dimensions, on a spatial `volume'
of linear size $R$ and at temperature $T=1/L$;
or, alternatively, as a statistical mechanics system living on a
two-space-dimensional lattice with a lattice spacing of
order $\sigma^{-1/2}$ (which has nothing to do with
the lattice on which the gauge theory may
or may not be regulated).

The field living on the worldsheet $(0\leq z_0\leq L,
0\leq z_1\leq R)$ is a bosonic field $\vec h$
with $D-2$ components.
Since it represents the displacement
of the worldsheet from the classical configuration, it has
engineering dimension of length.
The Polyakov loops are the propagators of static quarks,
so we are discussing the open-string case, and this is
reflected in the effective theory by imposing Dirichlet
boundary conditions on $\vec h$ at $z_1=0$ and $z_1=R$.
The action of the effective theory is
\ba
S &= & \sigma RL + S_{ \rm fluc}  \la{eq:Sexpansion} \\
S_{ \rm fluc}&=& \frac{\sigma}{2}\int d^2z \left[(\partial_\mu \vec h)^2
  +  c_1 (\partial_1 \vec h)^2 (\delta(R)+\delta(0))\right.\nn
&&+ \left. c_2^{(1)}(\partial_\mu \vec h)^2  (\partial_\nu \vec h)^2
  +   c_2^{(2)}(\partial_\mu \vec h \cdot\partial_\nu \vec h )^2
  + \dots \right]
\la{eq:gauss_action}
\ea
The Gaussian action
gives the leading contribution to the string spectrum at large $R$.
Each higher-order term is multiplied by an unknown dimensionless constant
$c_k^{(i)}$; some of them vanish/ are constrained by the open-closed
string duality~\cite{lw04}. The lower index on the action terms
above indicate how many more derivatives of $\vec h$ they contain than
$S_0$. On general grounds the correction to $V(R)$ associated 
with an operator of dimension 
$d_{\rm op}$ (when $\vec h$ is canonically normalised)
is $\frac{1}{R}(1/\sigma R^2)^{d_{\rm op}/2-1}$.
The parameter $\sigma$ in \eq\ref{eq:gauss_action} 
controls the amplitude of the fluctuations
around the classical solution; we assume that it is the same quantity as the one
appearing in \eq\ref{eq:Sexpansion}, which determines the classical energy. 
The fact that we are to treat it as a large 
quantity (it is the UV cutoff) clearly 
shows that we are doing a semi-classical calculation.

Lattice simulations~\cite{lw02,Meyer:2004hv}
have confirmed the leading, universal correction~\cite{luscher81} to the static 
potential predicted by the effective theory (\ref{eq:Sexpansion}, \ref{eq:gauss_action}):
\be
V(R)=\sigma R -\frac{\pi d_\perp}{24R}+O\left(\frac{1}{\sigma R^3}\right),
\ee
where $d_\perp$ is the number of (infinite)
transverse space-time dimensions. We now ask ourselves, what happens
if the latter are finite and periodic?

\section{Finite transverse dimensions to the string}
The first observation is that $V(R,L_\perp)$ is linearly confining $\forall~ L_\perp=(L_1,L_2)$.
The qualitative behaviour of the string tension is known in a few limits: 
 $\sigma(L_1,L_2)\propto 1/(L_1L_2)$ when $L_1,L_2\to 0$; 
 $\sigma(L_1) \propto 1/L_1^2$ when  $L_1\to 0$; and
 $\sigma(L_1)=\sigma(\infty)+O(e^{-m_GL_1})$ when $L_1$ is large, where
$m_G$ is the lightest glueball mass.
Secondly, $\sigma=\sigma(L_\perp)$ and $c_k=c_k(L_\perp)$ become
`arbitrary' functions of $\Lp$ (they can be computed on the lattice).
Since the effective theory cannot predict the values of these parameters,
we focus on the static potential in reduced units:
\be
v(r)\equiv V(R)/\sqrt{\sigma}=r-\frac{\pi d_\perp}{24r}+\dots,
 \quad r\equiv \sqrt{\sigma}R.
\ee

The question that now arises is, 
how does the L\"uscher correction vary with $\Lp$?
When a transverse dimension shrinks to zero,
the $D$-dimensional gauge theory should dimensionally reduce to a
$(D-1)$-dimensional one, where the L\"uscher correction
is given by $-\pi (d_\perp-1)/24r$. 
From the point of view of the effective string theory,
the coefficient is nothing but the central charge of the
2d-conformal theory describing the fluctuations of the worldsheet;
since it is given by the number of massless bosonic
degrees of freedom living on the worldsheet, it cannot vary continuously.
\emph{We thus expect the L\"uscher correction to have a discontinuity
at a phase transition of the worldsheet quantum field theory.}

Consider for simplicity the case of 3d SU($N$) gauge theory, with the size of the
(unique) transverse dimension given by $L_\perp$.
A possible interpretation is that  $L_\perp=1/T$ and we are looking at
the screening mass of a heavy quark at finite temperature.
What is now the effective string theory?

Due to the geometry of the target space,
the fluctuation field is parametrised by an \emph{angular variable}:
\be
h(z) = \frac{L_\perp \theta(z)}{2\pi}~ \Rightarrow ~
S_0 = \frac{K}{2}\int d^2 z (\partial_\mu \theta)^2, \quad
K = \frac{\sigma(L_\perp)L_\perp^2}{(2\pi)^2}.\nonumber
\la{eq:Z}
\ee
A low-energy effective theory with UV cutoff $\sqrt{\sigma}$
is equivalent to a lattice field theory with lattice spacing 
$O(\sigma^{-1/2})$~\cite{symanzik}. 
Two-dimensional systems with compact dynamical variables 
defined on a lattice are well-known in statistical mechanics. 
The most famous instance is the   XY-model:
\be
S_{\rm xy} = K_{\rm xy} \sum_{x,\mu}[1-\cos(\theta(x)-\theta(x+a\hat\mu))]
\ee
However, any model with the same symmetries is just as acceptable,
as long as we focus on the universal properties of this class of models.
Next, we review the properties of the XY-model that are relevant to
the dynamics of the QCD string.
%%%%%%%%%%%%%%%%%%%%%%%%%%%%%%%%%%%%%%%%%%%
\section{The XY-model as an effective description}
%%%%%%%%%%%%%%%%%%%%%%%%%%%%%%%%%%%%%%%%%%%
\begin{figure}
\begin{center}
\vspace{-0.5cm}
\includegraphics[width=7cm,angle=0]{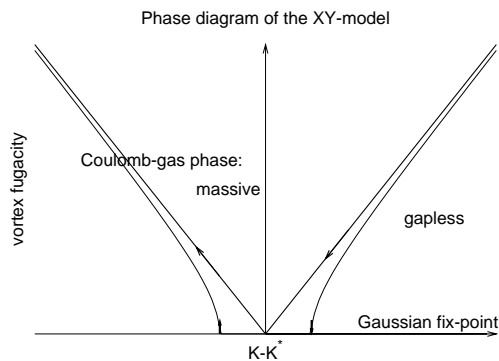} 
\end{center}
\caption{The phase diagram of the XY model.}
\la{fig:xy-phase-diag}
\end{figure}
At large $K_{\rm xy}$, the partition function is dominated by 
Gaussian fluctuations around the trivial $\theta=0$  vacuum.
However, the system admits \emph{vortex--antivortex}  configurations
 \be
 e^{i\theta(z)} = e^{i\theta_\infty}~\frac{z-z^+}{|z-z^+|}~
                  \frac{|z-z^-|}{z-z^-}
\la{eq:vort}
 \ee
whose energy is $\sim\log(R/a)$, where $R=|z^+-z^-|$.  At low 
temperatures, such configurations with $R$ of physical size are suppressed;
however their entropy is also proportional to $\sim\log(R/a)$, so that 
they drive a phase transition (of infinite order) at some finite $K_{\rm xy}^*\simeq2/\pi$,
as shown by Kosterlitz and Thouless~\cite{kt}. Clearly they have a strong 
disordering effect on the field $\theta$: a screening
mass is generated at small  $K_{\rm xy}$.
 
\begin{figure}
\vspace{1.2cm}
\hspace{4cm}
\begin{minipage}[l]{7cm}
\begin{picture}(0,0)(90,160)
\put(0,0){\line(1,0){200}}
\put(0,90){\line(1,0){200}}
\put(0,180){\line(1,0){200}}
\put(60,72){\circle*{4}}
\put(53,66){${\rm v}$}
\put(120,108){\circle*{4}}
\put(121,108){$\bar{\rm v}$}

\put(-25,4){$t=-\infty$}
\put(-25,94){$t=0$}
\put(-25,184){$t=+\infty$}

\put(0,-10){$0$}
\put(196,-10){$R$}

 \thicklines{
 \put(0,0){\line(0,1){180}}
 \put(204,0){\line(0,1){180}}

 \put(0,0){\vector(1,0){14}}
 \put(30,0){\vector(3,1){13}}
 \put(50,0){\vector(3,-1){13}}
 \put(90,0){\vector(3,-1){13}}
 \put(120,0){\vector(3,1){14}}
 \put(150,0){\vector(3,1){14}}
 \put(190,0){\vector(1,0){14}}

 \put(0,90){\vector(1,0){14}}
 \put(30,90){\vector(3,-1){12}}
 \put(50,90){\vector(1,-2){5}}
 \put(90,90){\vector(-1,0){14}}
 \put(120,90){\vector(0,1){14}}
 \put(150,90){\vector(1,1){10}}
 \put(190,90){\vector(1,0){14}}

 \put(0,180){\vector(1,0){14}}
 \put(30,180){\vector(3,-1){13}}
 \put(50,180){\vector(3,-1){13}}
 \put(90,180){\vector(3,1){13}}
 \put(120,180){\vector(3,1){14}}
 \put(150,180){\vector(3,-1){14}}
 \put(190,180){\vector(1,0){14}}

 }
\end{picture}
%\end{minipage}
%%%%%%%%%%%%%%%%%%%%%%%%%%%%%%%%%%%%%%%%%%%%%%%%%%%%%%%%%%%%%
\hspace{8cm}
%\begin{minipage}[r]{7cm}
%\vspace{2.5cm}
\begin{picture}(0,0)(50,190)
\put(-45,29){{\huge$\Leftrightarrow$}}
\put(-45,119){{\huge$\Leftrightarrow$}}
\put(-45,209){{\huge$\Leftrightarrow$}}
 \put(0,0){\line(1,0){50}}
 \put(0,20){\line(1,0){50}}
 \put(0,40){\line(1,0){50}}
 \put(0,60){\line(1,0){50}}

 \put(0,90){\line(1,0){50}}
 \put(0,110){\line(1,0){50}}
 \put(0,130){\line(1,0){50}}
 \put(0,150){\line(1,0){50}}

 \put(0,180){\line(1,0){50}}
 \put(0,200){\line(1,0){50}}
 \put(0,220){\line(1,0){50}}
 \put(0,240){\line(1,0){50}}

\put(5,25){\circle*{4}}
\put(-2,29){${\rm q}$}
\put(45,25){\circle*{4}}
\put(46,29){$\bar{\rm q}$}
\put(45,45){\circle*{4}}
\put(45,5){\circle*{4}}

\put(5,115){\circle*{4}}
\put(-2,119){${\rm q}$}
\put(45,95){\circle*{4}}
\put(46,99){$\bar{\rm q}$}
\put(45,135){\circle*{4}}
\put(45,115){\circle*{4}}

\put(5,205){\circle*{4}}
\put(-2,209){${\rm q}$}
\put(45,205){\circle*{4}}
\put(46,209){$\bar{\rm q}$}
\put(45,185){\circle*{4}}
\put(45,225){\circle*{4}}

 \thicklines{
 \put(5,25){\line(1,0){40}}
 \put(5,115){\line(2,-1){40}}
 \put(5,205){\line(1,0){40}}

\put(60,29){$t=-\infty$}
\put(60,119){$t=0$}
\put(60,209){$t=+\infty$}
}
\end{picture}
\end{minipage}
\vspace{6.5cm}
\caption{Physical interpretation of the vortices: the left-hand side,
where $\theta$ is represented as a unit vector $(\cos\theta,\sin\theta)$,
illustrates that the appearance of a vortex at time $t$ `followed' by an 
antivortex at time $\bar t$ induces one unit of clockwise winding 
of the variable $\theta(t_s,x)$ as  $x$ varies from $0$ to $R$
on a timeslice at time $t_s$ located between $t$ and $\bar t$.
Right, the corresponding interpretation for the  
string on the covering space of the torus: the string has non-trivial 
winding number along the transverse dimension in the interval
$[t,\bar t]$ and trivial winding number elsewhere.
} 
\label{fig:winding}
\end{figure}
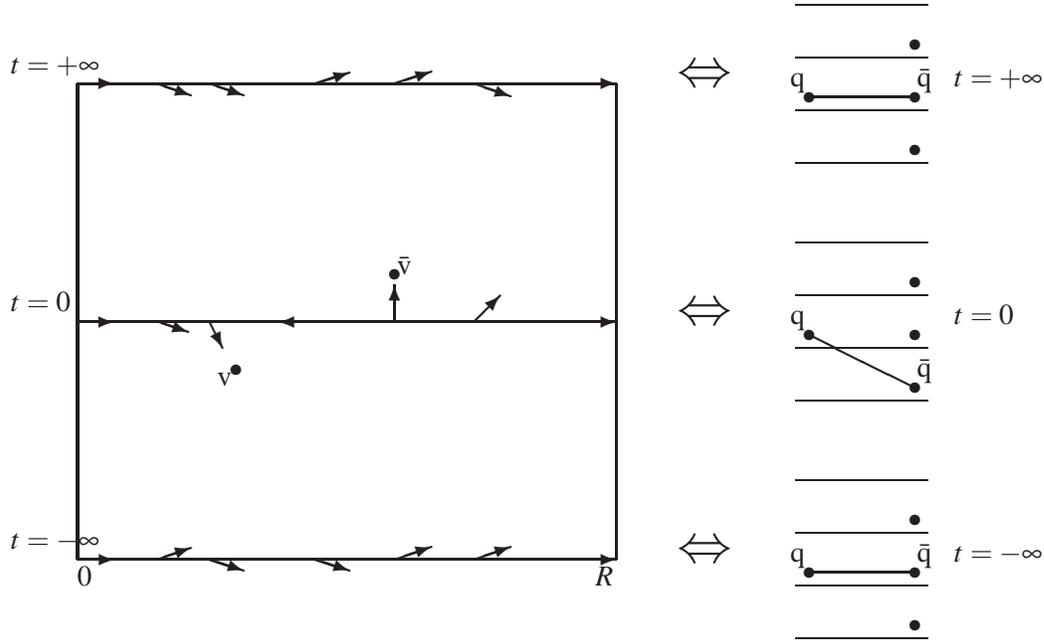

Let us apply these known facts to the effective theory on the string worldsheet.
In doing so, it is helpful to have in mind the following physical
interpretation of vortex configurations:
the action~(\ref{eq:Z}) is a saddle-point
expansion around the classical string solution which
connects a quark to an antiquark separated by distance $R$.
Clearly, there are also classical solutions
with a net winding number
around the compact transverse dimension. They have energy
\be
E_{\rm cl}(n)= \sigma(\Lp) \sqrt{R^2+n^2\Lp^2}\simeq
E_{\rm cl}(n=0)+ \frac{\sigma(\Lp) \Lp^2n^2}{2R}, \quad R\gg \Lp.
\ee
%The $n=1$ classical solution becomes as light as the
%first Gaussian excitation of the  $n=0$ solution when
%$\sigma(\Lp) \Lp^2=2\pi$. 
The different classical configurations
can only be connected if the string goes through more energetic
configurations. Therefore such transitions are classically forbidden
tunnelling processes. Quantum mechanically, one expects the tunnelling
to become a frequent fluctuation when the gap $E(1)-E(0)$ 
is of same order as the Casimir energy of the $n=0$ vacuum.
In this context, the worldsheet direction of size $R$ is
interpreted as the space-direction, and that of size $L$ as
the Euclidean time direction in a path integral treatment
of the quantum-mechanical string.
A vortex being a point-like object on the worldsheet,
it corresponds to a \emph{process} of the string describing
a transition from a state at $t=-\infty$ to another state
at $t=+\infty$. As long as we are considering the asymptotic
limit $L\to\infty$, only vortex-antivortex configurations
have a finite free energy in the 2d QFT. 
It is easy to see from \eq\ref{eq:vort} that a worldsheet containing a
${\rm v}\bar{\rm v}$ pair describes the `life' of a string that had
winding number $n=0$ at $t=-\infty$, goes through a state with
winding number $\pm1$, and returns to $n=0$ classical state (see
Fig.~\ref{fig:winding}).

Qualitatively, we can predict the following scenario for $v(r)$ as 
$\Lp$ is varied~\cite{Meyer:2005px}: at large $L_\perp$,
the periodicity plays no role for the
long-distance properties of the string fluctuations; the
Casimir energy still is $-\frac{\pi}{24R}$.
As $L_\perp$ decreases,  vortices affect the non-universal
      contributions to the energy levels. At 
 $L_\perp^* \simeq \sqrt{\frac{8\pi}{\sigma(\Lp^*)}}$
     (the precise value is non-universal),
      the KT phase transition takes place on the worldsheet.
For  $L_\perp<L_\perp^*$, the fluctuations acquire a mass gap
     and the Casimir energy is of order $e^{-2mR}$.

We can be more quantitative. 
Since vortices have a scaling dimension
$  x_v = \pi \kappa $ in the Gaussian approximation
$S_0 = \frac{\kappa}{2}\int d^2z(\partial_\mu \theta)^2$,
their contribution (in pairs) to energy levels at $L_\perp>L_\perp^*$ is 
\be
\Delta E(L_\perp,R)=\frac{c}{R}(\sigma(L_\perp)R^2)^{1-\pi\kappa}.
\ee
with $\kappa = \frac{\sigma(L_\perp)L_\perp^2}{(2\pi)^2} + O(1)$.

If we approach the phase transition from the $L_\perp<L_\perp^*$ side,
the correlation length diverges faster than any power law~\cite{kt}:
\be
m\sim m_0 e^{ -b\sqrt{K^*/(K^*-K)}},\quad {\rm Casimir~energy}\sim e^{-2mR}.
\la{eq:scaling}
\ee
By the same token, for a worldsheet of large but finite size, 
the phase transition becomes a crossover of spread 
$\delta K/ K^* =  b^2 / \log^2 m_0R $.

We have discussed the dynamics of a single transverse
dimension, since we were dealing with a three-dimensional
gauge theory. If we now move to the four-dimensional case,
where both components of the worldsheet fluctuation
field $\vec h$ are periodic, 
the key point to note is that the two components are
decoupled in the quadratic, renormalisable part of the 
action~(\ref{eq:gauss_action}). Therefore the universal
contributions to the string energy levels just add up.
Also the existence of topological configurations in one component
is unaffected by the presence of the other component.

\section{Conclusion}
We have studied an effective string theory for the
flux-tube in SU($N$) gauge theories
in three and four compact dimensions.
Let us summarise the outcome of our investigation of
the Polyakov loop correlator when one transverse dimension
is periodic and of size $\Lp$.
The periodic component of the fluctuation field is described
by an action of the XY-model type. It follows that
\bi
\item there is a phase transition of the worldsheet
      field theory at $\Lp^*=O(\sigma^{-1/2})$ where the
      periodic component acquires a mass gap and hence the central
      charge drops by one unit;
\item it is a Kosterlitz-Thouless phase transition:
      approaching $\Lp^*$ from below, the mass gap $m(\Lp)$ goes to zero
      in a universal way in terms of $K\equiv \sigma(\Lp)\Lp^2/4\pi^2$
      as given by Eq.~\ref{eq:scaling}. Once $\sigma(\Lp)$ is independently
       determined,
      we have a prediction for the functional form of $m(\Lp)$.
\item in the large $\Lp$ phase, there are (strongly suppressed)
      corrections to the string
      energy levels associated with the periodicity;
      they come in powers of $1/R$ which vary continuously and increase
      monotonously with $\Lp$; these corrections
      are exactly of order $1/R^3$ at $\Lp=\Lp^*$.
\ei

The gauge theory of course has the `deconfining' phase transition at
$\Lp\doteq 1/T_c=O(\sigma^{-1/2})$, and it would be intriguing if
$\Lp^*$ actually coincided with $1/T_c$. This can be tested in lattice simulations.
Numerical evidence was already presented in~\cite{Caselle:1993cb} showing that in the
three-dimensional $Z(2)$ gauge model $\Lp^*$ lies within  $10\%$ of $1/T_c$.


\begin{thebibliography}{99}
\bibitem{kuti_talk}
J.~Kuti, Plenary talk at this conference.
%%%%%%%%%%%%%%%%%%%%%%%%%%%%%%%%%%%%%%%%%%%%%%%%%%%%%%%%%%%%%%%%%%%%%%%%%%%%%
%\cite{Meyer:2005px}
\bibitem{Meyer:2005px}
H.~B.~Meyer,
%``Vortices on the worldsheet of the QCD string,''
arXiv:hep-th/0506034.
%%CITATION = HEP-TH 0506034;%%
%%%%%%%%%%%%%%%%%%%%%%%%%%%%%%%%%%%%%%%%%%%%%%%%%%%%%%%%%%%%%%%%%%%%%%%%%%%%%
%\cite{Caselle:1993cb}
\bibitem{Caselle:1993cb}
M.~Caselle, F.~Gliozzi and S.~Vinti,
%``On the relation between the width of the flux tube and T(c)**1 in lattice
%gauge theories,''
Nucl.\ Phys.\ Proc.\ Suppl.\  {\bf 34} (1994) 263
[arXiv:hep-lat/9403022];
%%CITATION = HEP-LAT 9403022;%%
%\cite{Caselle:1993mt}
%\bibitem{Caselle:1993mt}
M.~Caselle, R.~Fiore, F.~Gliozzi, P.~Guaita and S.~Vinti,
%``On the behavior of spatial Wilson loops in the high temperature phase of
%LGT,''
Nucl.\ Phys.\ B {\bf 422} (1994) 397
[arXiv:hep-lat/9312056].
%%CITATION = HEP-LAT 9312056;%%
 %%%%%%%%%%%%%%%%%%%%%%%%%%%%%%%%%%%%%%%%%%%%%%%%%%%%%%%%%%%%%%%%%%%%%%%%%%%%
%\cite{Luscher:2004ib}
\bibitem{lw04}
M.~Luscher and P.~Weisz,
%``String excitation energies in SU(N) gauge theories beyond the free-string
%approximation,''
JHEP {\bf 0407} (2004) 014
[arXiv:hep-th/0406205].
%%CITATION = HEP-TH 0406205;%%
%%%%%%%%%%%%%%%%%%%%%%%%%%%%%%%%%%%%%%%%%%%%%%%%%%%%%%%%%%%%%%%%%%%%%%%%%%%%
\bibitem{luscher81}
M.~Luscher, K.~Symanzik and P.~Weisz,
%``Anomalies Of The Free Loop Wave Equation In The Wkb Approximation,''
Nucl.\ Phys.\ B {\bf 173} (1980) 365;
M.~Luscher,
%``Symmetry Breaking Aspects Of The Roughening Transition In Gauge Theories,''
Nucl.\ Phys.\ B {\bf 180} (1981) 317.
%%%%%%%%%%%%%%%%%%%%%%%%%%%%%%%%%%%%%%%%%%%%%%%%%%%%%%%%%%%%%%%%%%%%%%%%%%%%
%\cite{Symanzik:1983dc}
\bibitem{symanzik}
K.~Symanzik,
%``Continuum Limit And Improved Action In Lattice Theories. 1. Principles And
%Phi**4 Theory,''
Nucl.\ Phys.\ B {\bf 226} (1983) 187.
%%CITATION = NUPHA,B226,187;%%
%%%%%%%%%%%%%%%%%%%%%%%%%%%%%%%%%%%%%%%%%%%%%%%%%%%%%%%%%%%%%%%%%%%%%%%%%%%%
%\cite{Luscher:2002qv}
\bibitem{lw02}
M.~Luscher and P.~Weisz,
%``Quark confinement and the bosonic string,''
JHEP {\bf 0207} (2002) 049
[arXiv:hep-lat/0207003];
%%%%%%%%%%%%%%%%%%%%%%%%%%%%%%%%%%%%%%%%%%%%%%%%%%%%%%%%%%%%%%%%%%%%%%%%%%%%
%\cite{Juge:2004xr}
%\bibitem{kuti}
K.~J.~Juge, J.~Kuti and C.~Morningstar,
%``QCD string formation and the Casimir energy,''
arXiv:hep-lat/0401032;
%%CITATION = HEP-LAT 0401032;%%
%\cite{Juge:2003sz}
%\bibitem{Juge:2003sz}
% K.~J.~Juge, J.~Kuti and C.~Morningstar,
%``The Casimir energy paradox of the QCD string,''
Nucl.\ Phys.\ Proc.\ Suppl.\  {\bf 129} (2004) 686
[arXiv:hep-lat/0310039];
%%CITATION = HEP-LAT 0310039;%%
%%%%%%%%%%%%%%%%%%%%%%%%%%%%%%%%%%%%%%%%%%%%%%%%%%%%%%%%%%%%%%%%%%%%%%%%%%%%
%%CITATION = HEP-LAT 0207003;%%
%\cite{Caselle:2005xy}
%\bibitem{Caselle:2005xy}
M.~Caselle, M.~Hasenbusch and M.~Panero,
%``Comparing the Nambu-Goto string with LGT results,''
JHEP {\bf 0503} (2005) 026
[arXiv:hep-lat/0501027].
%%CITATION = HEP-LAT 0501027;%%
%%%%%%%%%%%%%%%%%%%%%%%%%%%%%%%%%%%%%%%%%%%%%%%%%%%%%%%%%%%%%%%%%%%%%%%%%%%%
 %\cite{Kosterlitz:1973xp}
\bibitem{kt}
J.~M.~Kosterlitz and D.~J.~Thouless,
%``Ordering, Metastability And Phase Transitions In Two-Dimensional  Systems,''
J.\ Phys.\ CC {\bf 6}, 1181 (1973).
%%CITATION = JPCBA,C6,1181;%%
%%%%%%%%%%%%%%%%%%%%%%%%%%%%%%%%%%%%%%%%%%%%%%%%%%%%%%%%%%%%%%%%%%%%%%%%%%%%
%\cite{Meyer:2004hv}
\bibitem{Meyer:2004hv}
H.~B.~Meyer and M.~Teper,
%``Confinement and the effective string theory in SU(N $\to$ infinity): A
%lattice study,''
JHEP {\bf 0412} (2004) 031
[arXiv:hep-lat/0411039].
%%CITATION = HEP-LAT 0411039;%%
%%%%%%%%%%%%%%%%%%%%%%%%%%%%%%%%%%%%%%%%%%%%%%%%%%%%%%%%%%%%%%%%%%%%%%%%%%%%%%


\end{thebibliography}
\end{document}